\documentclass[amsmath,amssymb,showpacs,draft,preprint,floatfix]{revtex4}

\usepackage{graphicx}
\usepackage{latexsym}
\begin{document}
%\draft
\title{A family of exact eigenstates for a single trapped ion interacting with a
laser field}
% and for the Jaynes-Cummings model with counter-rotating terms}
\author{H. Moya-Cessa${}^{1,2}$,
D. Jonathan${}^{1,3}$ and P.L. Knight${}^{1}$}
\affiliation{${}^{1}$ QOLS,
Blackett Laboratory, Imperial College, London SW7 2BW, U.K. \\
${}^{2}$ INAOE, Coordinaci\'on de Optica, Apdo. Postal 51 y 216,
72000 Puebla, Pue., Mexico \\ ${}^{3}$  DAMTP, Centre for
Mathematical Sciences, University of Cambridge, Wilberforce Road,
Cambridge CB3 0WA, U.K. \\}
\date{\today}

\begin{abstract}

We show that, under certain combinations of the parameters
governing the interaction of a harmonically trapped ion with a
laser beam, it is possible to find one or more exact eigenstates
of the Hamiltonian, with no approximations except the optical
rotating-wave approximation. These are related via a unitary
equivalence to exact eigenstates of the full Jaynes-Cummings model
(including counter-rotating terms) supplemented by a static
driving term.
\end{abstract}

 \maketitle

\section{Introduction}
% Why ions matter
In recent years, trapped ions interacting with laser beams
\cite{wine0,wiem} have become an extremely interesting system for
the investigation of fundamental physics. For instance, they have
been used to generate nonclassical motional states
\cite{cira1,vogel,monr}, and have been proposed for applications
such as quantum computation
\cite{wine0,cira0,Steane00b,Jonathan00a} and precision
spectroscopy \cite{wine}.

% Ion trap not solved
Despite the relative simplicity of this system, the full
theoretical treatment of its dynamics is a nontrivial problem, as
the laser-ion interaction is highly nonlinear. Even in the
simplest case where only a single ion is in the trap, one is
usually forced to employ physically motivated approximations in
order to find a solution. A well-known example is the {\it
Lamb-Dicke} approximation \cite{LD}, in which the ion is
considered to be confined within a region much smaller than the
laser wavelength. Many treatments also assume a {\it weak
coupling} approximation, i.e., a sufficiently weak laser-ion
coupling constant. Under these conditions, tuning the laser
frequency to integer multiples of the trap frequency results in
effective Hamiltonians of the Jaynes-Cummings type \cite{JCM}, in
which the centre-of-mass of the trapped ion plays the role of the
field mode in cavity QED \cite{Blockley93a,Vogel95a}.

%Hector's transformation simplifies things
Recently, a new approach to this problem has been suggested
\cite{MCVBD99}, based on the application of a unitary
transformation $\hat{T}$ which linearises the total ion-laser
Hamiltonian. Under this transformation the Hamiltonian becomes
{\it exactly} (not effectively!) equivalent to the full
Jaynes-Cummings model (JCM), including counter-rotating terms,
together with an extra atomic driving term. Remarkably, the
ion-trap system is thus formally equivalent to an atom interacting
with a single-mode quantised electromagnetic field.

The existence of this correspondence is very useful, since it
allows us to map interesting properties of each model onto their
counterparts in the other. For example, we can make use of the
well-known `rotating-wave' limit where the JCM is analytically,
albeit approximately, soluble \cite{JCM}. Using the map
$\hat{T}^{\dagger}$ to translate this solution back into the
ion-trap scenario has led to the identification of a new
dynamically interesting regime for that system \cite{MCVBD99},
where phenomena such as ``super-revivals'' in the ion-laser
interaction can occur. The same solution has also led to a scheme
for realising substantially faster logic gates for quantum
information processing in a linear ion chain \cite{Jonathan00a}.

%What we do
In the present note, we apply this line of reasoning to the
problem of obtaining {\em exact} eigenstates of either system. To
our best knowledge, this has never been achieved in the case of
the ion-trap Hamiltonian. In the case of the JCM (with no extra
driving), exact eigenstates have been derived for those
combinations of system parameters at which level crossings occur
in the spectrum \cite{Reik82,Kus85}. In this paper we use a simple
physical ansatz similar to the one in \cite{Kus85} to derive a
family of exact eigenstates for the ion-trap system (again, these
are valid only under certain combinations of system parameters).
In the special case where the ion is driven on resonance, these
states correspond, via the mapping $\hat{T}$, to the ones found in
refs. \cite{Reik82,Kus85}. In addition, we study the asymptotic
limit of large Lamb-Dicke parameter $\eta$ in the ion-trap system,
which under $\hat{T}$ corresponds to a large coupling constant in
the JCM. We obtain in this way a very simple explanation of the
asymptotic forms of {\em all} eigenvalues and eigenvectors in this
limit (these have also been derived in \cite{Kus85} using other
methods).

\section{Ion trap-JCM correspondence}
%Revise Hector's approach
Let us start then by recalling the equivalence between the
ion-trap system and the JCM \cite{MCVBD99}. The Hamiltonian for
the ion-laser dipole interaction, with no approximations (except
the optical rotating-wave approximation \cite{wine}) can be
written as

\begin{equation}
H_{ion}=\nu \hat{n}+\frac{\delta }{2}\sigma _{z}+ \Omega \left(
\sigma_{+} \hat{D}(i\eta)+\sigma_{-}
\hat{D}^{\dagger}(i\eta)\right), \label{ion1}
\end{equation}
where $\hat{D}(i\alpha) = e^{i\alpha (a+a^{\dagger})}$ is the
displacement operator, $\nu $ is the harmonic trapping frequency,
$\delta =\omega _{atom}-\omega _{laser}$ the laser-ion detuning,
$\Omega $ the (real) Rabi frequency of the ion-laser coupling,
$\eta $ the Lamb-Dicke parameter, and we have chosen units where
$\hbar=1$. On the other hand, the Jaynes-Cummings Hamiltonian with
counter-rotating terms can be written as

\begin{equation}
H_{JCM}=\omega \hat{n}+\frac{\omega _{0}}{2}\sigma _{z}+i\lambda \left(
\sigma _{+}+\sigma _{-}\right) \left( a-a^{\dagger }\right).  \label{JCM1}
\end{equation}

Although these two models appear to be physically and
mathematically quite distinct, they are in fact exactly equivalent
The easiest way to see this is by rewriting eq.(\ref{ion1}) in a
notation where operators acting on the internal ionic levels are
represented explicitly in terms of their matrix elements:

\begin{equation}
H_{ion}=\left(
\begin{array}{cc}
\nu \hat{n}+\frac{\delta }{2} & \Omega \hat{D}\left( i\eta \right)
\\ \Omega \hat{D}^{\dagger }\left( i\eta \right) & \nu \hat{n}-\frac{\delta }{2}
\end{array}
\right) .  \label{ion2}
\end{equation}

Consider now the unitary operator
\begin{equation}
T=\frac{1}{\sqrt{2}}\left(
\begin{array}{ll}
\hat{D}^{\dag }\left( \beta\right) & \hat{D}\left( \beta\right) \\
-\hat{D}^{\dag }\left( \beta\right) & \hat{D}\left( \beta\right)
\end{array}
\right)\label{T}
\end{equation}
where $\beta = i\eta/2$. It is possible to check after some
algebra \cite{MCVBD99} that:
\begin{equation}
{\mathcal H} \equiv TH_{ion}T^{\dagger }=\left(
\begin{array}{cc}
\nu \hat{n}+\Omega +\frac{\nu \eta ^{2}}{4} & \frac{\iota \eta \nu }{2}%
\left( a-a^{\dag }\right) +\frac{\delta }{2} \\ \frac{\iota \eta
\nu }{2}\left( a-a^{\dag }\right) +\frac{\delta }{2} & \nu
\hat{n}-\Omega +\frac{\nu \eta ^{2}}{4}
\end{array}
\right)
\end{equation}
Returning to the usual notation, we obtain
\begin{equation}
{\mathcal H} =\nu \hat{n}+\Omega \sigma _{z}+\frac{\iota \eta \nu }{2}%
\left( \sigma _{+}+\sigma _{-}\right) \left( a-a^{\dag }\right) +\frac{%
\delta }{2}\left( \sigma _{+}+\sigma _{-}\right) +\frac{\nu \eta
^{2}}{4} \label{JCM2}
\end{equation}

Comparing with eq. (\ref{JCM1}) it can be seen that this is
precisely the Jaynes-Cummings interaction, supplemented by two
additional terms: the first corresponds to an extra static
electric field interacting with the atomic dipole, and the second
is just a constant energy shift. In particular, a purely
Jaynes-Cummings form is recovered when $ \delta =0$, corresponding
to a resonant laser-ion interaction in eq. (\ref {ion1}). Of
course, the various parameters of the Hamiltonian in eq.
(\ref{JCM2}) have different meanings than they do in eq.
(\ref{ion1}): $\nu $ becomes the cavity field frequency $\omega $,
$2\Omega $ the atomic transition frequency $\omega _{0}$, $\delta
$ the coupling strength with the static field and $\eta $ the
ratio between the Jaynes-Cummings Rabi frequency $2\lambda $ and
the cavity frequency $\omega $. In what follows we shall refer to
eq. (\ref{JCM2}) as the `Jaynes-Cummings picture' of the ion-trap
Hamiltonian, eq. (\ref{ion1}).

As noted above, this correspondence enables one to map interesting
properties of each model onto the other. In this paper, we will
use it to translate between corresponding eigenstates of each
system (it is clear that, if $\left| \psi \right\rangle $ is an
eigenstate of $H_{ion}$, then $ T\left| \psi \right\rangle $ is a
corresponding one for ${\cal H}$). In this regard it is important
to point out that, although $H_{ion}$ and ${\cal H}$ both describe
systems consisting of a two-level atom interacting with a bosonic
mode, one should not identify each of these subsystems with their
counterparts after the transformation has been applied. This is
due to the fact that $T$ is an {\it entangling } transformation:
separable internal-motional states of the trapped ion can be
mapped into entangled atom-cavity states in the corresponding
cavity QED system.

\section{Eigenstates in the asymptotic limit}\label{sec:asympt}

In order to demonstrate the convenience of this correspondence,
let us consider what happens to the systems above in the
`asymptotic' limit where $\eta \gg \Omega/\nu$ (this problem has
been studied in \cite{Kus85} in the special case $\delta=0$, using
different methods). In the ion-trap picture, this limit
corresponds to having a very large atomic recoil after the
absorption of a photon from the laser; in the JCM picture, to
having a very large atom-mode coupling constant. From the point of
view of current ion-trap and cavity QED experiments, neither of
these conditions can be considered realistic. Nevertheless, it may
well be possible to engineer such a situation artificially, for
example an `effective' JCM with large `coupling constant' can be
obtained when a single ion is simultaneously illuminated with weak
lasers tuned to the first red and blue sidebands \cite{wine0}.

It is simpler to consider the problem first in the JCM picture. If
$\eta \gg \Omega/\nu$, then the term containing $\Omega$ in eq.
(\ref{JCM2}) makes a negligible contribution to the total energy.
We can therefore expect that the asymptotic eigenvalues and
eigenvectors will not be affected if this term is in fact absent.
Setting $\Omega =0$ implies that $\sigma_x\equiv
\sigma_++\sigma_-$ becomes a constant of the motion, so the
asymptotic eigenstates can be chosen to be of the form
$\left|\pm\right\rangle\otimes \left|\phi_{\pm}\right\rangle$,
where $\left| \pm\right\rangle = \left(\left| g\right\rangle
\pm\left| e\right\rangle\right)/\sqrt{2}$. In order to determine
their form explicitly, and also their corresponding eigenvalues,
let us switch back to the ion-trap picture. Setting $\Omega =0$ in
eq. (\ref{ion1}) reduces the Hamiltonian to that of an {\em
uncoupled} spin-boson system. The asymptotic eigenstates of the
ion-trap are therefore simply $\left|e\; m\right\rangle, \left|g\;
m\right\rangle$. Transforming back to the JCM picture using eq.
(\ref{T}), we obtain thus $\left|\phi_{\pm}\right\rangle =
\left|\pm i\eta/2; n \right\rangle$, where $\left|\alpha
;k\right\rangle \equiv \hat{D}\left(\alpha \right) \left|
k\right\rangle $ is a displaced number state \cite{dispnum} (note
that they are independent of $\delta$). The corresponding
eigenvalues are $m\nu\pm\delta/2$  \cite{note2}. Note in
particular that, when $|\delta|=k\nu$ for integer $k$, all but $k$
of these energy levels are degenerate. In the special case
$\delta=0$ it can in fact be shown from symmetry considerations
alone that the entire asymptotic spectrum must be two-fold
degenerate \cite{Sandro}). Conversely, no such asymptotic
degeneracies occur for $|\delta|\neq k\nu$.

\section{Eigenstates from a simple ansatz}

Let us return now to the ion-trap Hamiltonian, eq. (\ref{ion1}).
We will construct an {\it ansatz }which allows the determination
of exact eigenstates of this system, provided certain relations
are satisfied between the parameters $\Omega ,\delta ,\eta $.
Consider the possibility of finding an eigenstate of the form

\begin{equation}
\left| \psi_{ion}^{m+} \right\rangle = \left| g\right\rangle
\left| \phi\right\rangle + \frac{\Omega }{\nu }\left| e
\right\rangle \sum_{n=0}^{m+1}c_{n}\left| n\right\rangle,\;\;\;
c_{m+1}. \neq 0,\label{ansatz1}
\end{equation}
Let us see whether the eigenvalue equation
\begin{equation}
H_{ion}\left| \psi_{ion}^{m+}\right\rangle =E_m^+\left|
\psi_{ion}^{m+}\right\rangle, \label{eigveq}
\end{equation}
can be satisfied. Eq. (\ref{ion2}) shows that this requires
$\left| \phi \right\rangle $ to be of the form
\begin{equation}
  \left| \phi \right\rangle = D^{\dag }\left( i\eta \right)\sum_{n=0}^{m+1}d_{n}
  \left| n\right\rangle  =
\sum_{n=0}^{m+1}d_n\left|-i\eta; n\right\rangle.
\end{equation}
We thus require
\begin{equation}
H_{ion}\left| \psi \right\rangle = {\displaystyle
\sum_{n=0}^{m+1}\Omega\left(\left(n+\frac{\delta}{2\nu}\right)c_n
+d_n\right)\left|e\right\rangle \left|n\right\rangle}+
{\displaystyle  \left( \frac{\Omega^2}{\nu}c_n +d_n\left( \hat{n}
-\frac{\delta}{2}\right)\right)\left|g\right\rangle\left| -i\eta
;n\right\rangle} . \label{eigveq1}
\end{equation}

Now, using the simple fact that $\hat{D}^{\dag
}\left(\alpha\right)\hat{a}\hat{D}\left(\alpha\right) = \hat{a}
+\alpha$ \cite{MandelandWolf}, it is easy to show that displaced
number states satisfy the recursion relation
\begin{equation}
\hat{n}\left| \alpha;k \right\rangle =(\left| \alpha \right| ^{2}+k)\left| \alpha;k
\right\rangle +\alpha \sqrt{k+1}\left| \alpha ;k+1\right\rangle
+\alpha^* \sqrt{k}\left| \alpha ;k-1\right\rangle .
\label{dispeq1}
\end{equation}
Substituting then eqs. (\ref{ansatz1}), (\ref{eigveq1}) and
(\ref{dispeq1}) into eq. (\ref{eigveq}) and using the fact that
$\left\{\left| \alpha;k \right\rangle\right\}_{k=0}^{\infty}$ is
an orthonormal basis gives the following eigenstate conditions:
\begin{equation}
E_m^+ = (m+1)\nu +\frac{\delta}{2},\;\;\;\;\;\;\;c_{n}=\left\{
\begin{array}{c}
 {\displaystyle \frac{d_{n}}{m+1-n}};\;\;0\leq n\leq m \\ \\
 {\displaystyle \frac{i\eta\nu ^{2}}{\Omega ^{2}}} \sqrt{m+1}\;d_{m};\;\;n=m+1
\end{array}
\right.
\end{equation}
where the $d_{n}$ coefficients satisfy $d_{m+1}=0$ and
\begin{equation}\label{cond1a}
\left[
\begin{array}{ccccc}
\varepsilon _{m} & -i\eta   &  &  &  \\ i\eta   & \varepsilon
_{m-1} & -i\eta  \sqrt{2} &  &  \\ & i\eta  \sqrt{2} & \ddots &
\ddots  &  \\ &  & \ddots  & \varepsilon _{1} & -i\eta \sqrt{m}
\\ &  &  & i\eta  \sqrt{m} & \varepsilon _{0}
\end{array}
\right] \left[
\begin{array}{c}
d_{0} \\
\\
\vdots  \\
\\
d_{m}
\end{array}
\right] =\vec{0}
\end{equation}
where
\begin{equation}\label{eps+}
\varepsilon _{j}= \left( j+1-\eta ^{2}\right) +\frac{\delta}{\nu}
-\frac{\Omega ^{2}}{ (j+1)\nu^2 }.
\end{equation}

Using operator $\hat{T}$ we can map this state into a
corresponding eigenstate of the generalised JCM model in eq.
(\ref{JCM2})
\begin{equation}
\left| \psi_{JCM}^{m+} \right\rangle = \hat{T}\left|
\psi_{ion}^{m+} \right\rangle = \sum_{n=0}^{m+1}\left(d_n\left| +
\right\rangle - \frac{\Omega }{\nu }c_n\left| -
\right\rangle\right)\left|-i\eta/2;n\right\rangle. \label{eigjcma}
\end{equation}
Once again we verify that the solution for this system can be
easily expressed in terms of a displaced number state basis.

Further solutions can also be found if we note that the
Hamiltonian $ H_{ion}$ is invariant under the combined
transformations
\begin{equation}\label{symms}
  \left| e\right\rangle \leftrightarrow \left| g\right\rangle
;\text{ }\delta \leftrightarrow -\delta ;\text{ }\eta
\leftrightarrow -\eta.
\end{equation}
Defining then
\begin{equation}
\left| \psi_{ion}^{m-} \right\rangle = \frac{\Omega }{\nu
}\sum_{n=0}^{m+1}c'_{n}\left| g\right\rangle\left| n\right\rangle
+\sum_{n=0}^{m}d_n^{\prime}\left| e\right\rangle \left| i\eta
,n\right\rangle,
\end{equation}
and applying this symmetry transformation to eqs. (\ref{eigveq1})-
(\ref{eps+}) we can see that $\left| \psi_{ion}^{m-}
\right\rangle$ is also an eigenstate of $H_{ion}$, with eigenvalue
$E_m^-=(m+1)\nu -\frac{\delta}{2}$, as long as
\begin{equation}
c'_{n}=\left\{
\begin{array}{c}
 {\displaystyle \frac{d_{n}^{'}}{m+1-n}};\;\;0\leq n\leq m \\ \\
 {\displaystyle -\frac{i\eta\nu ^{2}}{\Omega ^{2}}} \sqrt{m+1}\;d'_{m};\;\;n=m+1
\end{array}
\right.
\end{equation}
where the $d'_{n}$ coefficients satisfy an equation analogous to
eq.(\ref{cond1a}) but with
\begin{equation}\label{eps-}
\varepsilon _{j}= \left( j+1-\eta ^{2}\right) -\frac{\delta}{\nu}
-\frac{\Omega ^{2}}{ (j+1)\nu^2 }.
\end{equation}

Condition (\ref{cond1a}) means that each of these ans\"{a}tze
succeeds only for certain combinations of  $\Omega ,\delta ,\eta
$. This is because the tridiagonal matrix above (which we will
refer to as $M^{\pm}_m$) must have a zero eigenvalue, or
equivalently $\det M^{\pm}_m=0$. Since $\det M^{\pm}_m$ is a
polynomial of degree $m+1$ in $\Omega^2$, $\delta$ or $\eta^2$,
fixing two of these quantities determines up to $m+1$ real
solutions for the third one. For $m\leq3$ it is possible to solve
this `compatibility' condition algebraically, resulting in
explicit expressions in terms of any of the three parameters. For
example, in terms of $\eta$, the first few solutions are
\begin{eqnarray}\label{nodes}
  m=0&:&\;\; \eta^2 = 1 -a +sd\\
  m=1&:&\;\; \eta^2 = 2+sd-\frac{3a}{4}\pm\sqrt{\frac{a^2}{16}-\frac{a}{2}+2+sd}
\end{eqnarray}
where $a = \left(\Omega/\nu\right)^2$, $d = \delta/\nu$ and
$s=\pm1$ according to which ansatz we are referring to. For $m>3$
a numerical solution is necessary.

The physical meaning of these solutions can be clarified by
examining the spectrum of $H_{ion}$, obtained by numerically
diagonalising this operator (or, equivalently, $\cal{H}$). Let us
first examine the case where $\delta =0$ (previously studied in
\cite{Reik82,Kus85}). In fig. (\ref{JCM}a) we plot the first few
energy levels as functions of $\eta$, for the case where $\Omega =
\nu/2$ (corresponding to a resonant JCM with no extra driving in
eq.(\ref{JCM2})). It can be readily seen that pairs of adjacent
energy levels form braids around the lines $E=m\nu$ for integer
$m$, converging to these lines in the limit of large $\eta$ (as
expected from our considerations in section \ref{sec:asympt}). The
level crossings occur precisely on top of these lines, and there
are $m$ such degeneracy points on the line $E=m\nu$ \cite{note2}.
(In fact, it can be shown \cite{Kus85} that these are the {\em
only } energy values at which degeneracies may occur in this
system).

Note now that, when $\delta=0$ equations (\ref{eps+}) and
(\ref{eps-}) coincide, and thus, $\left| \psi_{ion}^{m\pm}
\right\rangle$ are simultaneous (in fact, {\em degenerate})
eigenstates for any combination of $\Omega, \eta$ at which  $\det
M^{\pm}_m=0$. Another way of understanding this fact is to note
that the ion-trap Hamiltonian $H_{ion}$ has in this case an extra
symmetry: it commutes with the parity-like observable $\sigma_{x}
\exp(i\pi a^{\dagger}a)$ (in the JCM picture, the corresponding
symmetry operator is $\sigma_z \exp(i\pi a^{\dagger}a)$). It can
be easily checked that neither $\left| \psi_{ion}^{m+}
\right\rangle$ nor $\left| \psi_{ion}^{m-} \right\rangle$ have
this symmetry, but simple linear combinations of them do. Every
solution to the equation $\det M^{\pm}_m=0$ must therefore
correspond to one of the level crossing points; in other words,
the location of these points can be calculated directly from this
condition (for $m>3$ this cannot be done algebraically, of
course). In fact, it turns out that, when $\Omega =\nu/2$, this
equation always has $m$ real and positive solutions for $\eta$
\cite{Kus85}, and so {\em every} crossing point in fig. \ref{JCM}a
is accounted for in this way. (See \cite{Kus85} for a detailed
discussion of what happens as the value of $\Omega$ is changed).

For more general values of $\delta$ the situation changes as
follows (fig \ref{JCM}b,c): whenever $\delta/\nu=k$ where $k$ is a
nonzero integer (i.e., the laser is tuned to a sideband
transition), it is easy to see that $E^+_m =
E^-_{m+k}=(m+1+k/2)\nu$. In other words $\left| \psi_{ion}^{m\;+}
\right\rangle$ and $\left| \psi_{ion}^{m+k\;-} \right\rangle$  are
again degenerate eigenstates, and again the degenerate energies
coincide with the asymptotic values for large $\eta$ (recall sec.
\ref{sec:asympt}). In fact, it can be verified that $\det
M^{+}_m=0\Rightarrow \det M^-_{m+k}=0$, so that both eigenstates
occur for the same system parameters, thus corresponding once
again to line crossings in the spectrum (fig. \ref{JCM}b). (For
low values of $m,k$ this can be shown explicitly, for higher
values one can again resort to numerical methods). It seems likely
that this coincidence may once again be due to an underlying
symmetry, however in this case we have not been able to determine
it. Finally, when $\delta$ is not an integer, the crossings in the
spectrum become avoided crossings (fig. \ref{JCM}c), and in fact
it is easy to show, using a method analogous to that in ref.
\cite{Kus85}, that {\em no} degeneracies can occur in the system.
In this case, the solutions obtained above simply mark (some of)
the points where the spectrum crosses its asymptotic value (fig.
\ref{JCM}c).

\section{Conclusion}

In conclusion, we have shown that in certain circumstances it is
possible to obtain exact eigenstates for the Hamiltonian of a
single trapped ion. These states are often (but not always)
connected with the existence of level crossings in the system's
energy spectrum viewed as a function of the Lamb-Dicke parameter.
Another property is that they are always expressible as a finite
expansion in terms of certain physically well-motivated
vibrational states (number states and displaced number states). In
this regard, it is worth recalling a long-standing conjecture
concerning a possible {\em complete} solution to the JCM in terms
of known functions (valid for any values of the system
parameters). It was suggested in \cite{Reik86a} that the
eigenstates of this system may always be expressible as a finite
expansion in terms of a certain basis set, which in Bargmann
representation corresponds to a particular class of transcendental
functions. As far as we know this proposal has never been
analytically verified, although numerically it seems to hold
\cite{russians,Reik86a}. In the light of our present results, it
is natural to speculate that this result should still hold in the
presence of an additional static field, and thus should also apply
to the single-ion system.

H. M.-C. would like to thank CONACYT (Consejo Nacional de Ciencia
y Tecnolog\'{\i}a) and the Royal Society of London for support,
and also Imperial College for hospitality. D.J. was supported in
part by the Brazilian Government's Conselho Nacional de
Desenvolvimento Cient\'{\i}fico e Tecnol\'{o}gico (CNPq) and by
Pembroke College, Cambridge.

\begin{figure}[hbt]
\caption{\label{JCM} First few energy levels (in units of $\hbar$)
of a single trapped ion driven by a laser beam, as a function of
the Lamb-Dicke parameter $\eta$, for coupling constant
$\Omega=\nu/2$. Dark circles denote parameter values for which
exact eigenstates of this system can be found using the ansatz
described in section IV. (a) When the laser is resonant
($\delta=0$), a solution exists whenever the energy $E$ is an
integer multiple of the trap frequency $\nu$, or in other words
whenever the curves intersect the horizontal asymptotes to which
all energy levels tend in the limit $\eta \rightarrow\infty$.
These are also the level crossing points in the spectrum
\cite{Kus85}. (b) A similar behaviour occurs when $\delta/\nu=k$
for integer $k\neq0$ (i.e., the laser is tuned to one of its
sidebands) (here $k=1$). In this case the level crossings are at
points where $E = (m+1+ 0.5)\nu$, which are again the asymptotic
values of $E$. (c) Finally, when $\delta$ is not tuned to a
sideband ($\delta = 0.5\nu$ here), level crossings become avoided
crossings, and our ansatz finds exact eigenstates only at some of
the points where the curves cross the asymptotes $E_m =
(m+1.25)\nu$.}

\end{figure}


\begin{thebibliography}{aaaa}
\bibitem{note2} Note that this is the spectrum of $\cal{H}$ in eq.
(\ref{ion2}), which includes a constant energy shift proportional
to $\eta^2$. In other words, the spectrum of the `regular' JCM of
eq. (\ref{JCM1}) braids around the curves $m-\eta^2/4$, as is
well-known.
\end{thebibliography}
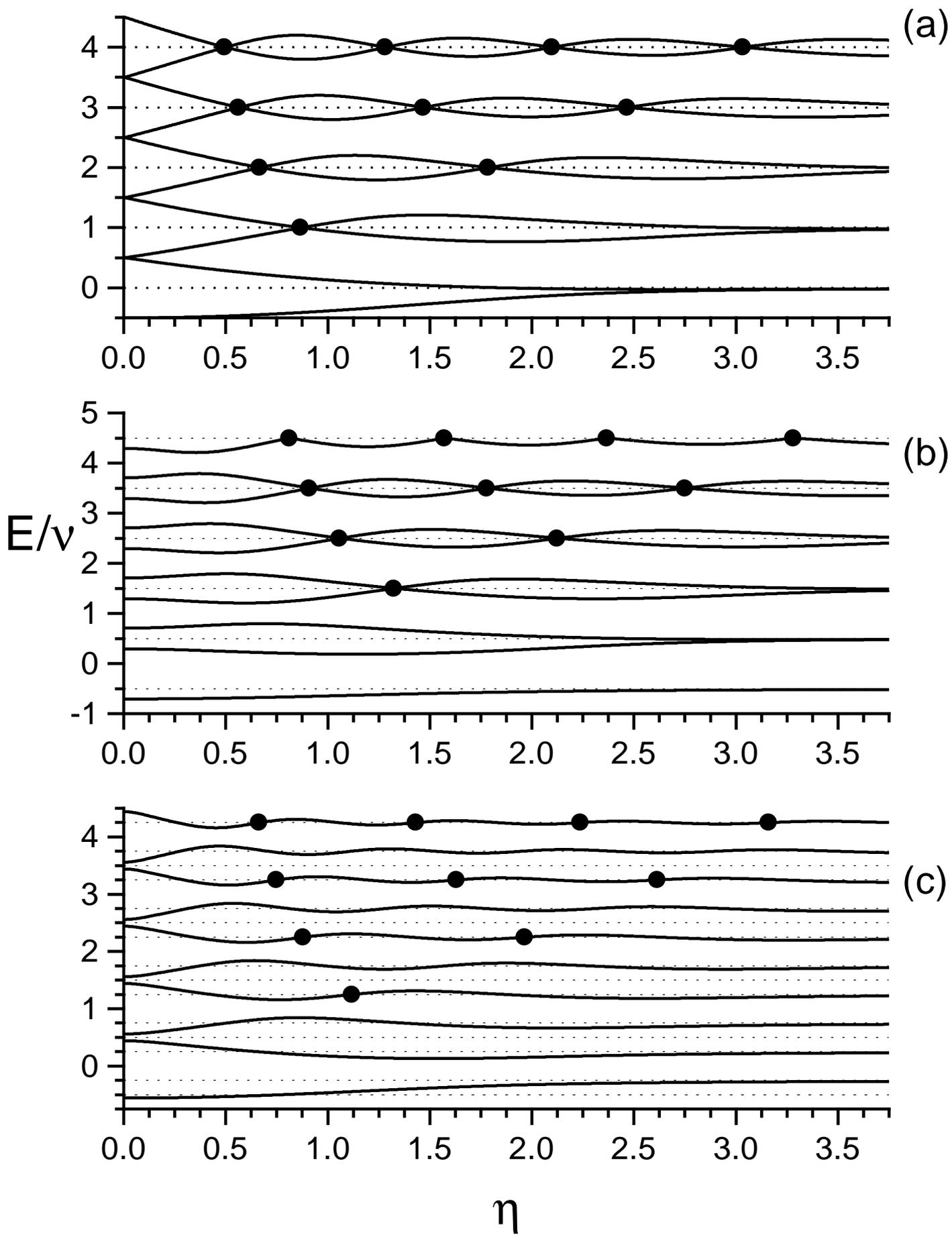
\end{document}